\documentclass[12pt]{article}
\usepackage{graphicx}

\usepackage[svgnames]{xcolor} 
\usepackage{subfig}
\usepackage{times} 

\graphicspath{{figures/}} 
\usepackage{booktabs} 
\usepackage[font=small,labelfont=bf]{caption} 
\usepackage{amsfonts, amsmath, amsthm, amssymb} 
\usepackage{wrapfig} 
\usepackage[labelformat=empty]{caption}
\usepackage{fontawesome}
\usepackage[utf8]{inputenc} 
\usepackage[T1]{fontenc}    
\usepackage{hyperref}       
\usepackage{url}            
\usepackage{booktabs}       
\usepackage{amsfonts}       
\usepackage{nicefrac}       
\usepackage{microtype}      
\usepackage{lipsum}
\usepackage{lineno}
\usepackage{graphicx}
\usepackage{subfig}
\usepackage{floatrow}
\usepackage{amsmath}

\newcommand\pubnumber{} 
\newcommand\pubdate{\today}

\def\institute{Regional Centre of Advanced Technologies and Materials, Joint Laboratory of
Optics of Palacký University and Institute of Physics AS CR, Faculty of Science,
Palacký University, 17. listopadu 12, 771 46 Olomouc, Czech Republic \\
petr.baron01@upol.cz}

\def\Title#1{\begin{center} {\Large #1 } \end{center}}
\def\Author#1{\begin{center}{ \sc #1} \end{center}}
\def\Address#1{\begin{center}{ \it #1} \end{center}}

\newcommand\pubblock{\rightline{\begin{tabular}{l} \pubnumber\\
         \pubdate  \end{tabular}}}
\newenvironment{Abstract}{\begin{quotation}  }{\end{quotation}}
\newenvironment{Presented}{\begin{quotation} \begin{center} 
             PRESENTED AT\end{center}\bigskip 
      \begin{center}\begin{large}}{\end{large}\end{center} \end{quotation}}
\def\Acknowledgements{\bigskip  \bigskip \begin{center} \begin{large}
             \bf ACKNOWLEDGEMENTS \end{large}\end{center}}

\def\beq{\begin{equation}}
\def\eeq#1{\label{#1}\end{equation}}
\def\eeqn{\end{equation}}

\def\beqa{\begin{eqnarray}}
\def\eeqa#1{\label{#1}\end{eqnarray}}
\def\eeqan{\end{eqnarray}}

\let\bar=\overbar

\def\Dslash{\not{\hbox{\kern-4pt $D$}}}
\def\dslash{\not{\hbox{\kern-2pt $\del$}}}

\def\msb{{\bar{\ssstyle M \kern -1pt S}}}

\begin{document}
\begin{titlepage}
\pubblock

\vfill
\Title{Fully Bayesian Unfolding with Regularization}
\vfill
\Author{\underline{P.~Baron}}
\Address{\institute}
\vfill
\begin{Abstract}
Fully Bayesian Unfolding differs from other unfolding methods by providing the full posterior probability of unfolded spectra for each bin. We extended the method for the feature of regularization which could be helpful for unfolding non-smooth, over-binned or generally non-standard shaped spectra. To decrease the computation time, the iteration process is presented. 
\end{Abstract}
\vfill
\begin{Presented}
$12^\mathrm{th}$ International Workshop on Top Quark Physics\\
Beijing, China, September 22--27, 2019
\end{Presented}
\vfill
\end{titlepage}
\def\thefootnote{\fnsymbol{footnote}}
\setcounter{footnote}{0}
\vspace{-12pt}
\section{Introduction}
\noindent
Unfolding is the process of correcting measured spectra for finite resolution and efficiency effects in high energy physics from the detector to the particle level which is an experiment-independent result. Currently, mostly used methods are the Bayes (D’Agostini) and SVD methods as implemented in the RooUnfold package \cite{Prosper:2011zz}. \par
This study aims to provide an automated usage of the modern Fully Bayesian Unfolding (FBU), improved by an automatic iteration over the parameters phase space to establish limits for faster convergence, to add implementation of regularization so far missing in available FBU implementations, and study appropriate sampling, \emph{e.g.} using No-U-Turn sampling in Hamiltonian Monte Carlo algorithm \cite{sampling}.\par 
The procedure is tested on spectra of top quarks in proton-proton collisions at \\$\sqrt{\text{s}}=14$ TeV generated by the MadGraph generator \cite{madgraph}, showered by Pythia8 \cite{pythia} to provide the particle level, and finally with the detector level simulated by Delphes \cite{delphes}. Conclusions relevant to the usage of FBU in current high energy experiments at LHC are drawn.\par
\vspace{-12pt}
\section{Fully Bayesian Unfolding}
The schematic equation of the unfolding can be written as
\begin{equation}
    p = \frac{1}{\epsilon}\cdot M^{-1}\cdot\eta\cdot(D - B);
\end{equation}
where $p$ is the unfolded particle-level spectrum, $M^{-1}$ is the symbol for a given unfolding method with the migration matrix $M$, $\epsilon$ and $\eta$ are efficiency and acceptance corrections respectively $D$ is data spectrum from which the background $B$ is subtracted.\par
The Fully Bayesian Unfolding method is based on the conditional probability and Bayes theorem. Its advantage compared to other methods is the possibility of choosing a prior probability called prior $\pi(T)$. $P(T|D)$ is then the probability density that the unfolded spectrum $T$ is inferred and can be written as
\begin{equation}
    P(T|D) = \frac{ P(D|T) \cdot \pi(T)}{\mathrm{Norm.}}
\end{equation}
using the given data $D$. $\mathrm{Norm.}$ is the normalization constant. Ideally, the unfolded spectrum $T$ is equal to particle spectrum $P$.\par 
The probability density $P(D|T)$ is proportional to the likelihood function $L(D|T)$ and the~prior
{\small
\begin{equation}
\left.\begin{aligned}
&P(T|D) \propto L(D|T) \cdot \pi(T) = \\&=  \left (  \prod_{i=1}^{n=\mathrm{bins}} \frac{1}{\epsilon_i} \frac{\left (\sum\limits_{j=1}^{n=\mathrm{bins}} M_{ij} T_j \right)^{\left [\eta_{i}(D_i-B_i)\right ]}}{\left [\eta_{i}(D_i-B_i)\right ]!} e^{-\left(\sum\limits_{j=1}^{n=\mathrm{bins}} M_{ij} T_j\right)} \right ) e^{-\tau S(T)}.
\end{aligned}\right.
\end{equation}
The practical formula is
\begin{equation}
\left.\begin{aligned}
&P(T|D)\sim \left ( \prod_{i=1}^{n=\mathrm{bins}}  \frac{1}{\epsilon_i} \frac{1}{\sqrt{2\pi \left(\sum\limits_{j=1}^{n=\mathrm{bins}} M_{ij} T_j\right)}}e^{-\left [\eta_{i}(D_i-B_i) - \left(\sum\limits_{j=1}^{n=\mathrm{bins}} M_{ij} T_j\right)\right ]^2} \right ) e^{-\tau S(T)};\\
\end{aligned}\right.
\label{maineq}
\end{equation}
}%
where $e^{-\tau S(T)}$ is the prior. If regularization strength parameter $\tau$ is zero then the prior is flat ($\pi(T)=1$). However, if $\tau$ is positive the regularization is applied according to the regularization function $S(T)$. The corrections
\begin{equation}
\epsilon = \frac{P_{\text{particle, proj. from M}}}{P_{\text{level}}} ;~
\eta = \frac{D_{\text{data, proj. from M}}}{\tilde{D}}\\
\end{equation}
are called efficiency and acceptance corrections respectively, 
where $P_{\text{particle, proj. from M}}$ and $D_{\text{data, proj. from M}}$  represent spectra obtained by making particle resp. detector level projections from the migration matrix. $P_{\text{level}}$ and $\tilde{D}$ are the original particle and detector level spectra taken from simulation.
\vspace{-12pt}
\subsection{Production of the test spectra}
For the test spectra the process of top anti-top quark production at 14 TeV in $\ell$+jets channel is simulated using Madgraph and Delphes. The resolved topology (a scenario when all the final state objects needed to reconstruct the top quark are reconstructed seperately in the detector) is studied. The detection of the particles was simulated using ATLAS Delphes card as a part of the package.

\begin{figure}[htbp]
\subfloat[]{\includegraphics[width=0.33\linewidth]{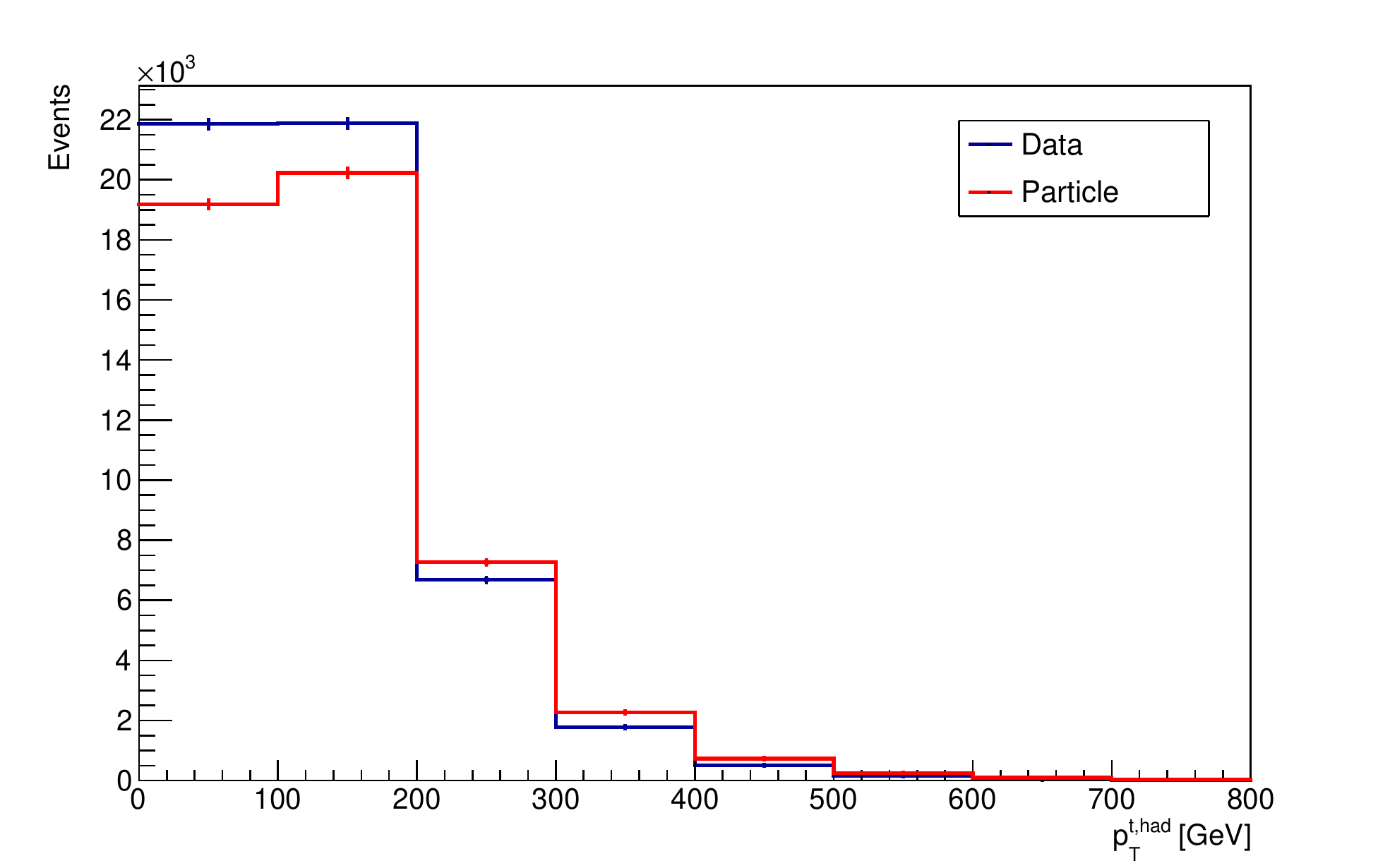}}
\subfloat[]{\includegraphics[width=0.33\linewidth]{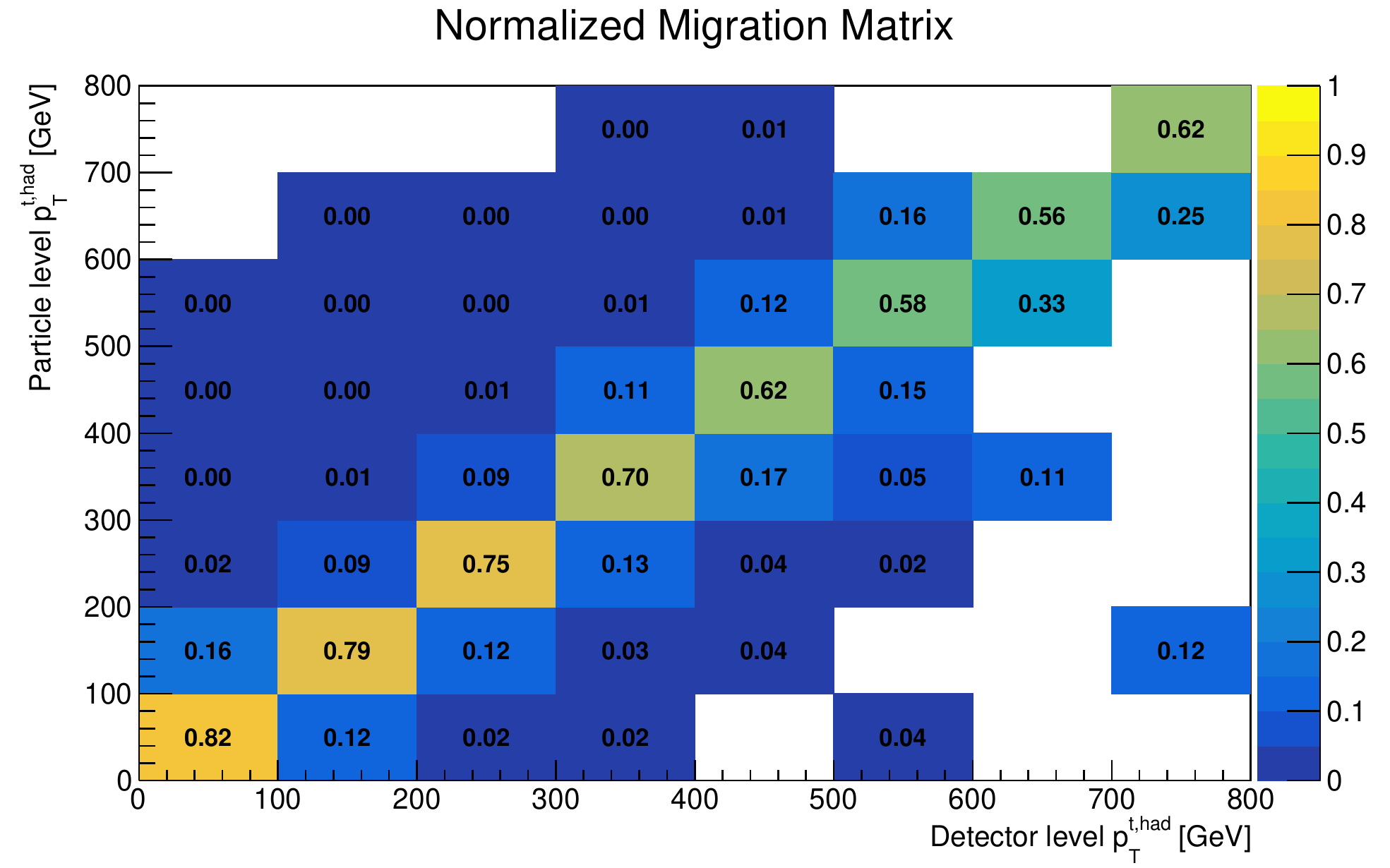}}
\subfloat[]{\includegraphics[width=0.33\linewidth]{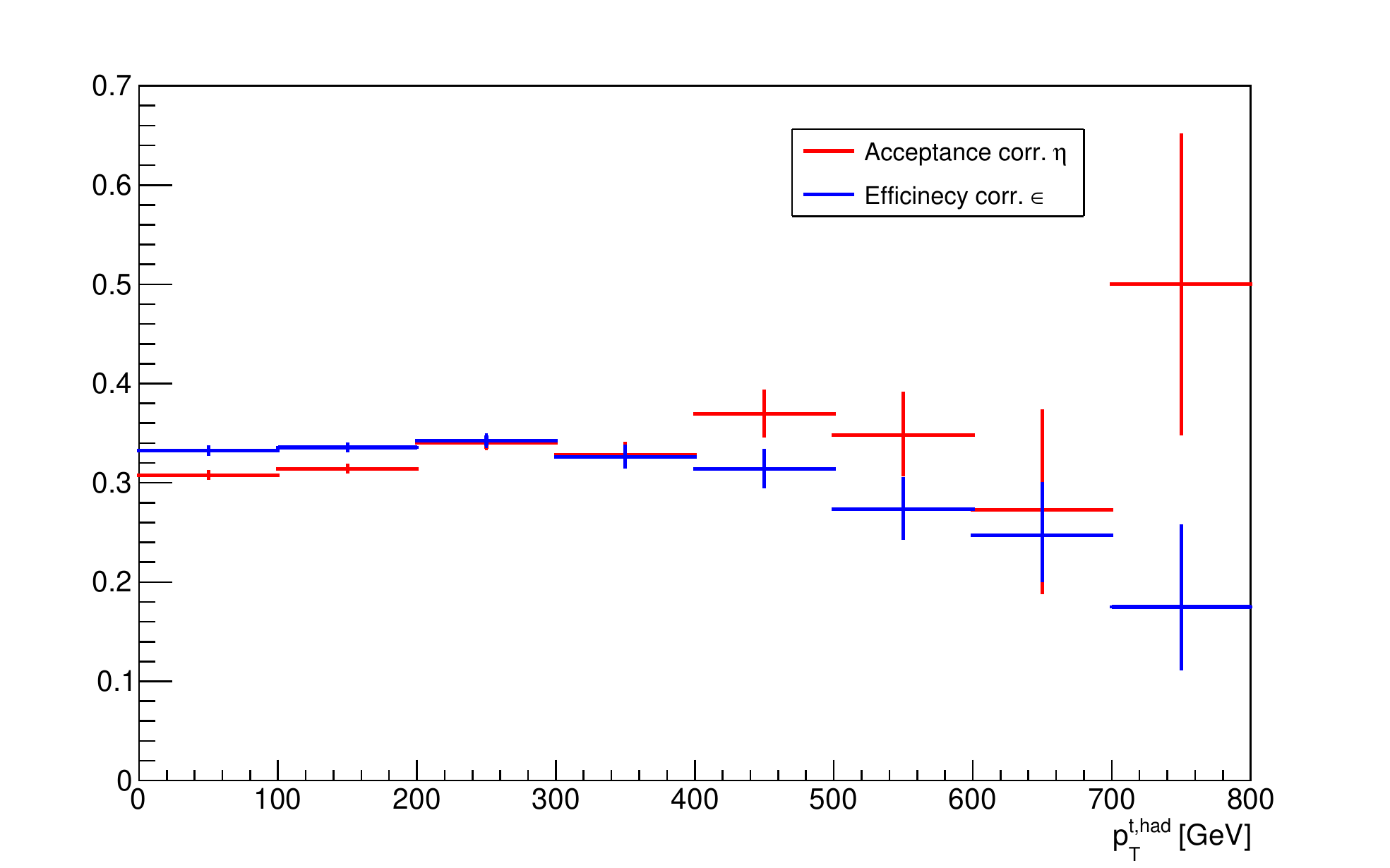}}
\caption{Figure 1: FBU ingredients. \textbf{a)} Detector-level (blue) and particle-level (red) spectra. \textbf{b)} Migration matrix between particle and detector levels. \textbf{c)} Efficiency (blue) and acceptance (red) corrections as a function of transverse momentum of hadronically decaying top quark.}
\label{fig:DriftCorrection}
\end{figure}
The unfolded spectrum is derived from posteriors which are calculated for each bin $i$ by marginalization
\begin{equation}
    p_i(T_i|D) = \int \int P(T|D) dT_1...dT_{i-1}dT_{i+1}...dT_{N}
\end{equation}
The unfolded spectra are taken as the fitted mean of the fit Gauss function and the uncertainty is taken as posterior $\sigma_{\text{gauss}}$ standard deviation.

\begin{figure}[h!]
\subfloat[]{\includegraphics[width=0.45\linewidth,height=0.25\linewidth]{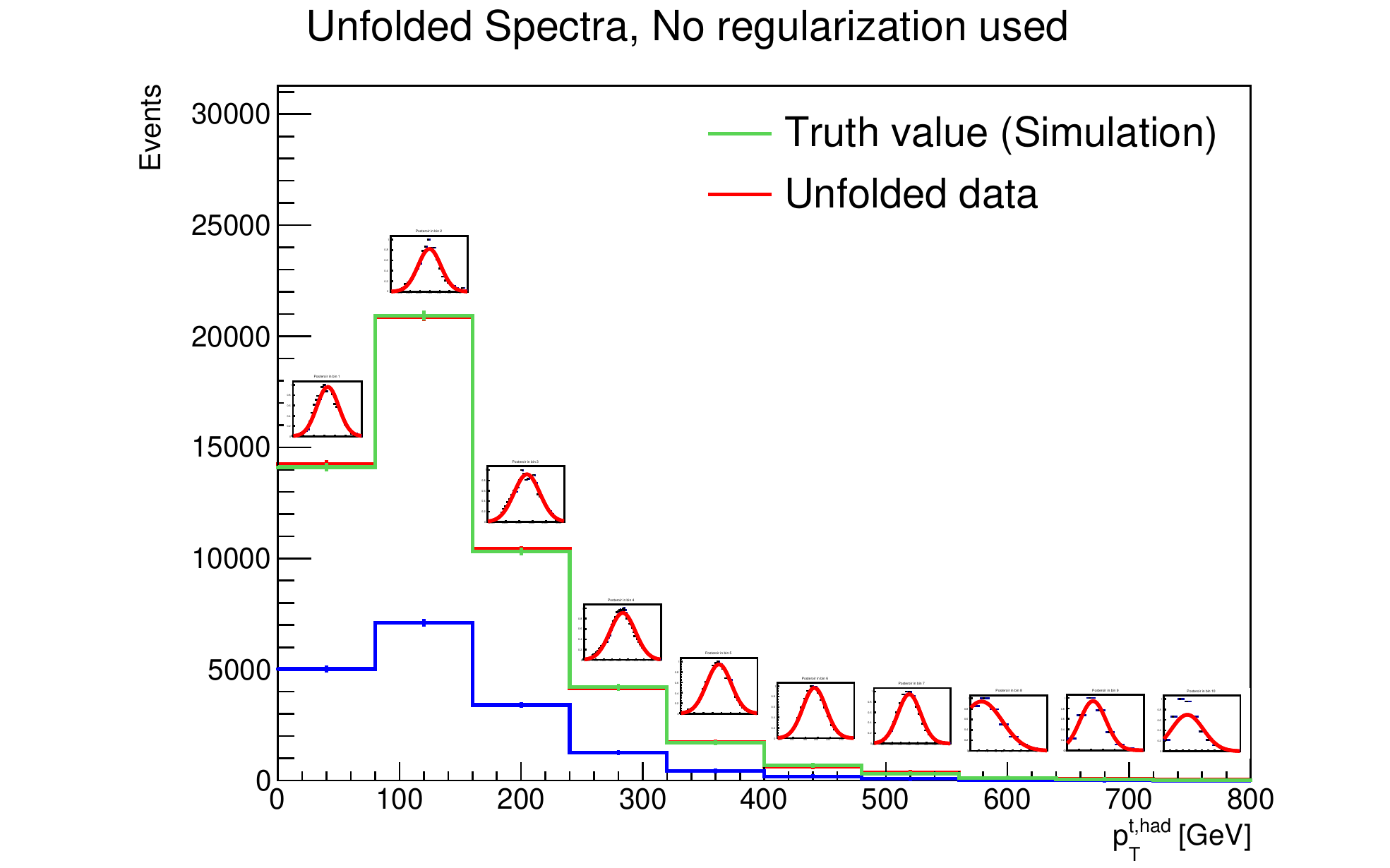}}
\subfloat[]{\includegraphics[width=0.45\linewidth,height=0.25\linewidth]{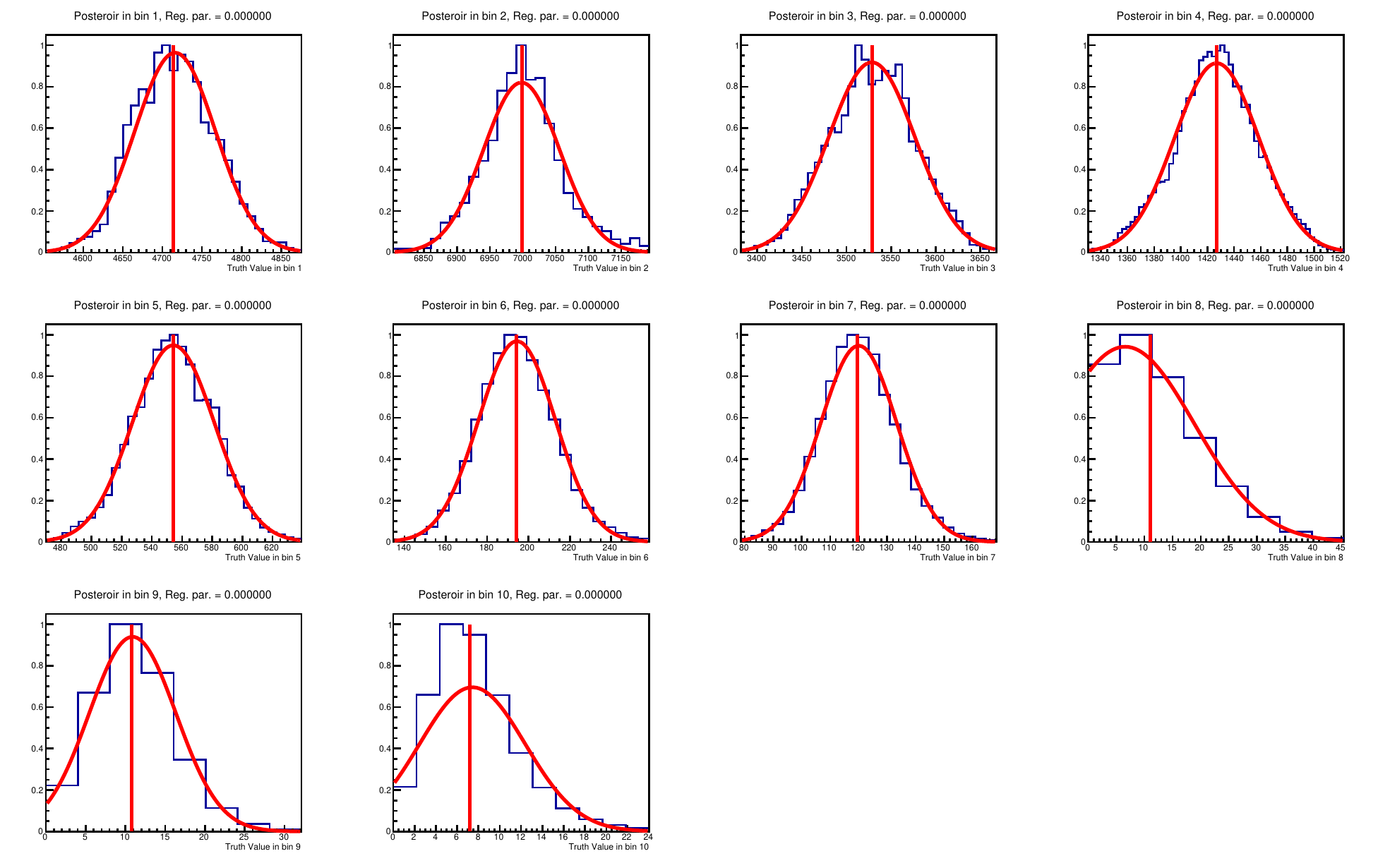}}
\caption{Figure 2: Unfolded spectrum of transverse momentum of hadronically decaying top quark (a) derived from the posteriors (b).}
\label{fig:poster}
\end{figure}

\vspace{-24pt}
\section{Phase space estimation and iterative FBU}
\begin{figure}[htbp]
\subfloat[]{\includegraphics[width=0.3\linewidth]{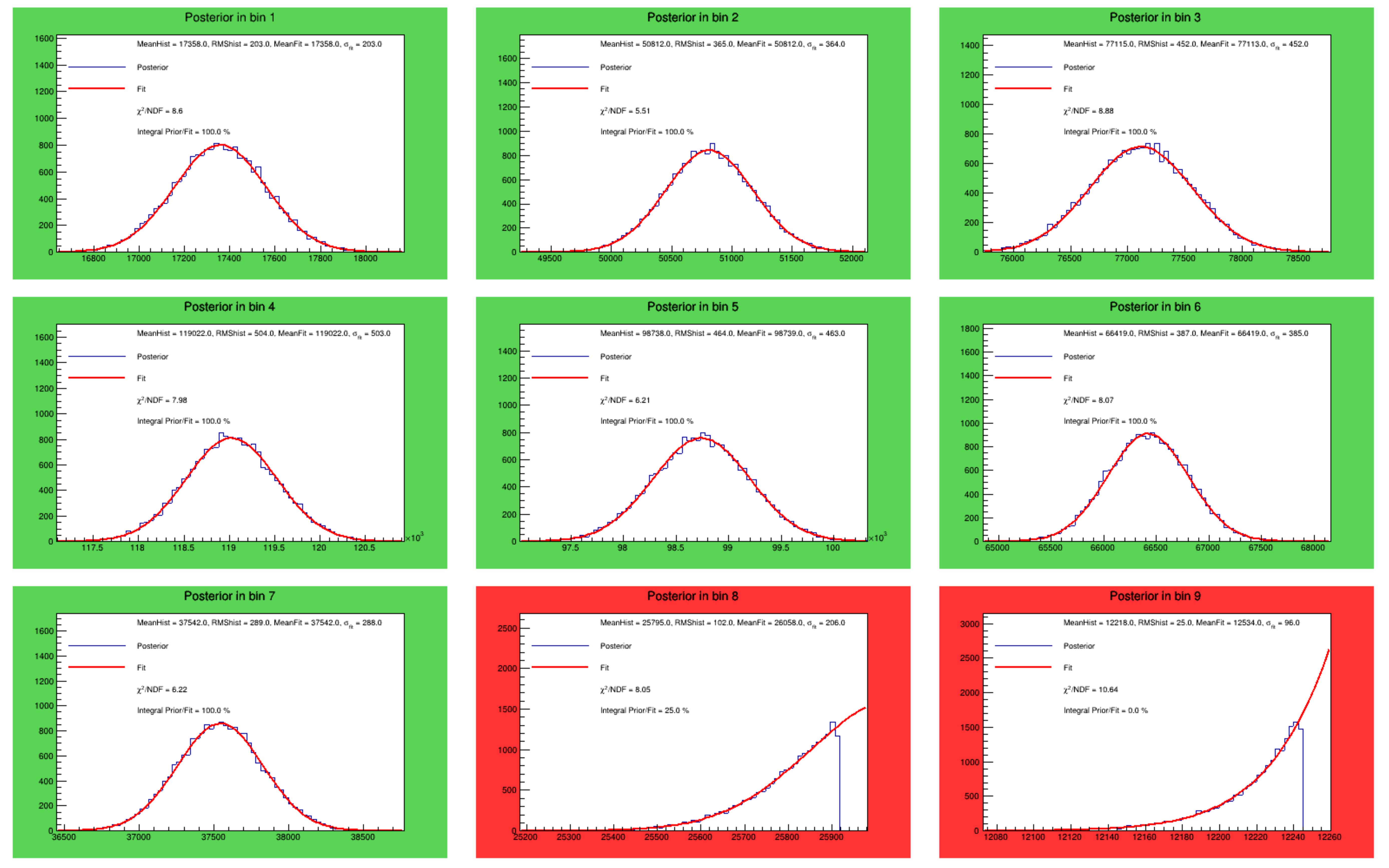}}
\hspace{1em}%
\subfloat[]{\includegraphics[width=0.3\linewidth]{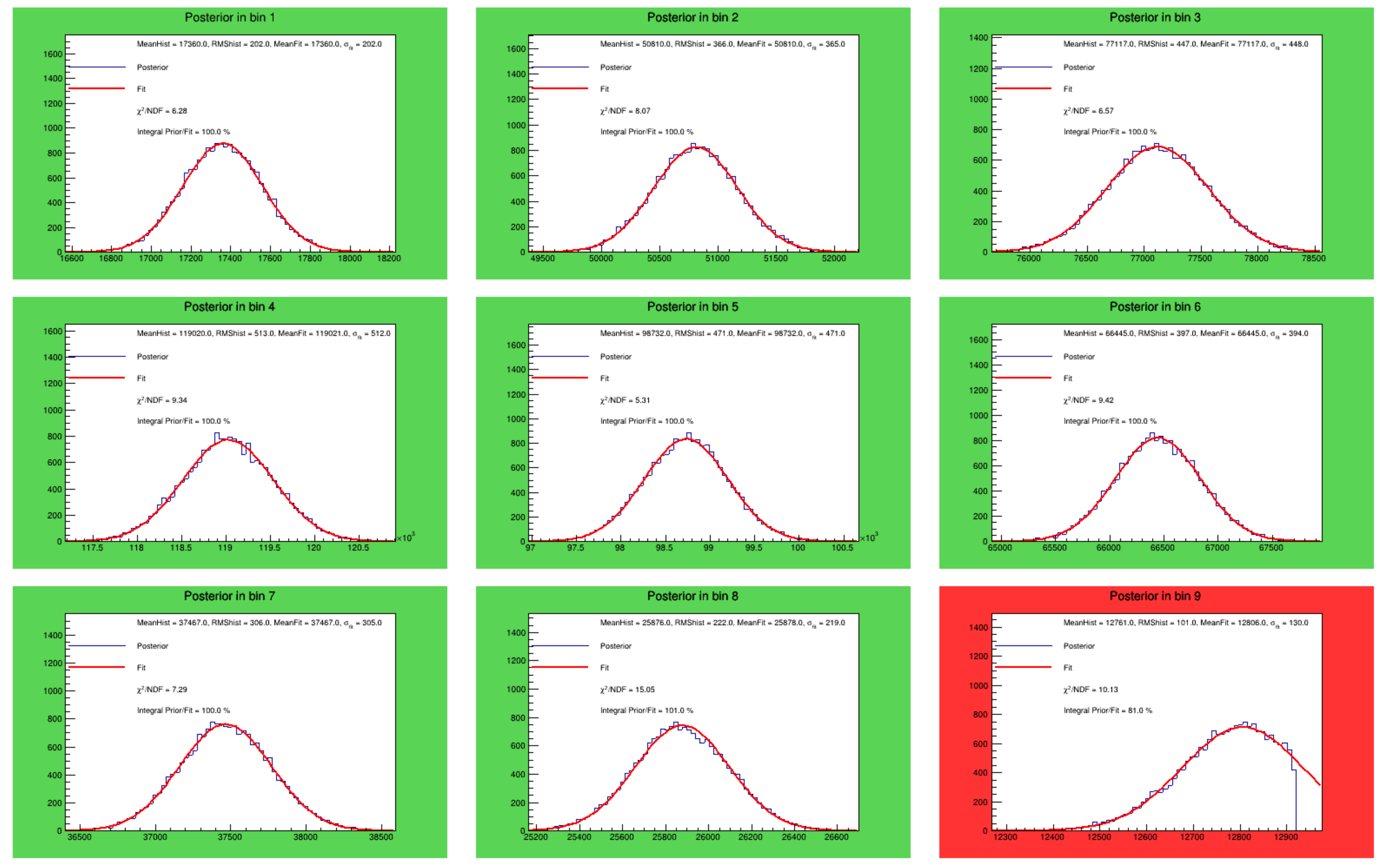}}
\hspace{1em}%
\subfloat[]{\includegraphics[width=0.3\linewidth]{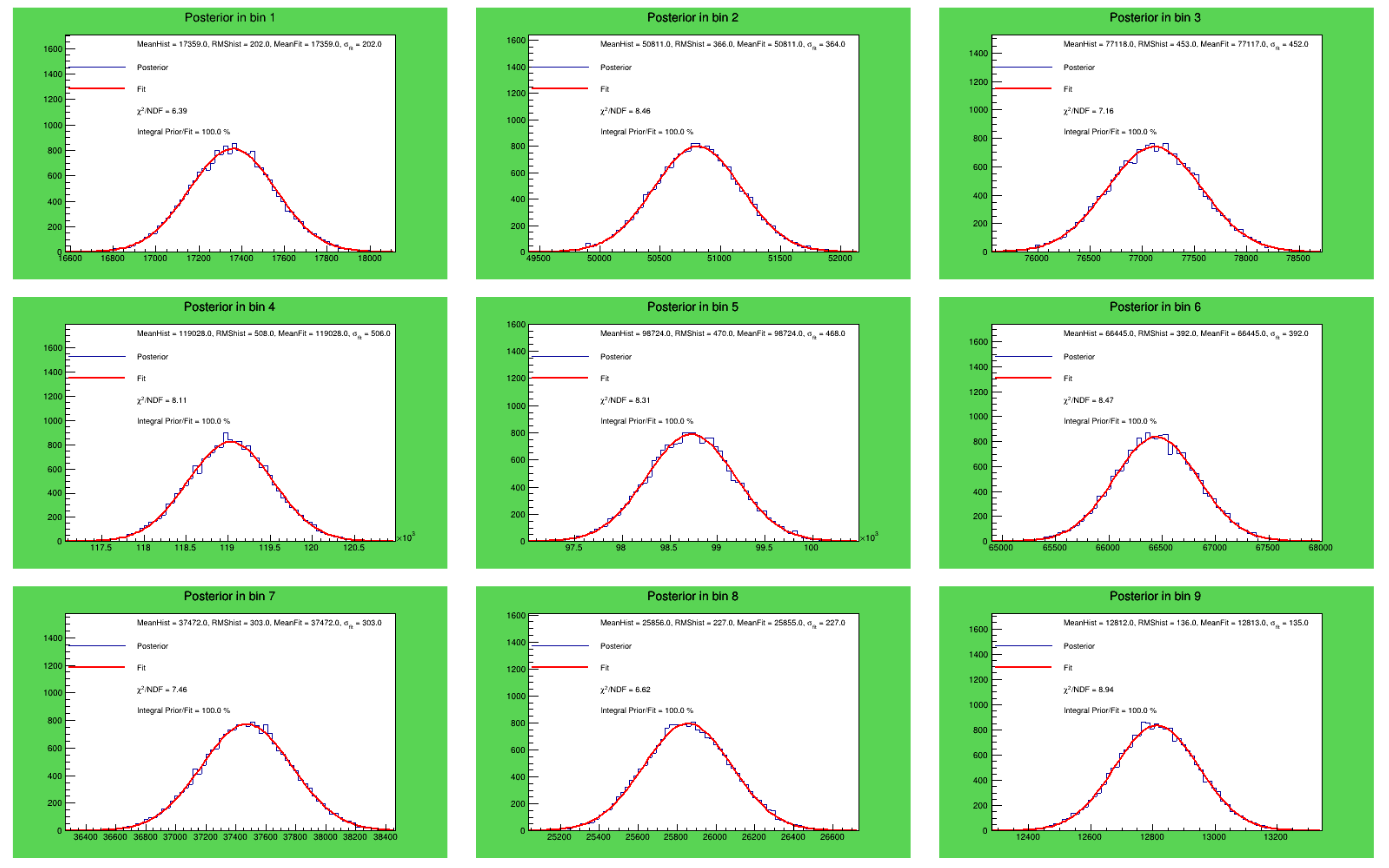}}
\caption{Figure 3: Estimation of the phase space of unfolding after a) one b) two and c) three iterations.}
\label{poster}
\end{figure}

In order to decrease the computation power the phase space limits need to be estimated. In this study the estimation of phase space for each $i$-th bin is given as
\begin{equation}
\small{
\left.\begin{aligned}
    &\left \{ \text{min}_i; \text{max}_i \right \} =\left \{ 0.01\left ( D_{i} \cdot \frac{D_{i~\text{data, proj. from M}}}{P_{i~\text{particle, proj. from M}}}\right ) ; 1.99\left (D_{i} \cdot \frac{D_{i~\text{data, proj. from M}}}{P_{i~\text{particle, proj. from M}}}\right)\right \}.
    \end{aligned}\right.
    }
\end{equation}
However this estimation does not need to be universal for all the bins which causes problem shown in the last two bins in Figure \ref{poster} (a). In this case the posteriors are fitted by a Gauss function $g(\mu, \sigma)$ and the unfolding algorithm is launched again with new phase space given as
\begin{equation}
    \left \{ \text{min}_i; \text{max}_i \right \} = \left \{ \mu_i - 4 \cdot \sigma_i~;~ \mu_i + 4 \cdot \sigma_i \right\};
\end{equation}
where $\mu$ and $\sigma$ stands for the mean and one standard deviation of the Gauss function obtained by fitting the posteriors. Figures \ref{poster} (b) and \ref{poster} (c) show the result of two more iterations.

\vspace{-24pt}
\section{FBU with regularization}
The regularization in the Fully Bayesian Unfolding can be introduced in a very natural way and is represented by the prior $\pi(T) = e^{-\tau S(T)}$. In this study the regularization function $S(T)$ is chosen as the curvature of the truth pseudo experiment $T$
\begin{equation}
    S(T) = \sum\limits_{t=2}^{N-1}(\Delta_{t+1,t}-\Delta_{t,t-1})^2;
\end{equation}
where
\begin{equation}
    \Delta_{t_1,t_2} = T_{t_1}-T_{t_2}
\end{equation}
\emph{i.e.} using the sum of second derivatives of the truth spectrum. The unfolding equation then reads
\begin{equation}
\small{
    \left.\begin{aligned}
&P(T|D) \propto L(D|T) \cdot \pi(T) \sim
\\&\sim  \left ( \prod_{i=1}^{n=\mathrm{bins}} \frac{1}{\epsilon_i} \frac{1}{\sqrt{2\pi \left(\sum\limits_{j=1}^{n=\mathrm{bins}} M_{ij} T_j\right)}}e^{-\left [\eta_{i}(D_i-B_i) - \left(\sum\limits_{j=1}^{n=\mathrm{bins}} M_{ij} T_j\right)\right ]^2} \cdot e^{-\tau \cdot \sum\limits_{t=2}^{N-1}(\Delta_{t+1,t}-\Delta_{t,t-1})^2}\right ).
\end{aligned}\right.
}
\end{equation}
However, in order to save computation power, $\text{log}(P(T|D))$ is computed and for the task of finding the maximum the likelihood is converted into a hypothetical movement of a free particle in the  $n$-dimensional ($n$ = number of bins) space with potential given by the likelihood. This method is called Hamiltonian Monte Carlo \cite{sampling}.\par
To reduce the computation time, the mapping of the phase space in our code is implemented using Leapfrog and BuildTree functions described in~\cite{sampling}.\par
The regularization might be useful in cases of non-smooth, over-binned or generally non-standard shaped spectra. In this study the pseudorapidity spectrum of the top quark pair was chosen due to its non-trivial shape double peak structure to show possible applications.
\vspace{-12pt}
\section{Results}
\begin{figure}[h!]
\subfloat[]
{\includegraphics[width=0.45\linewidth]{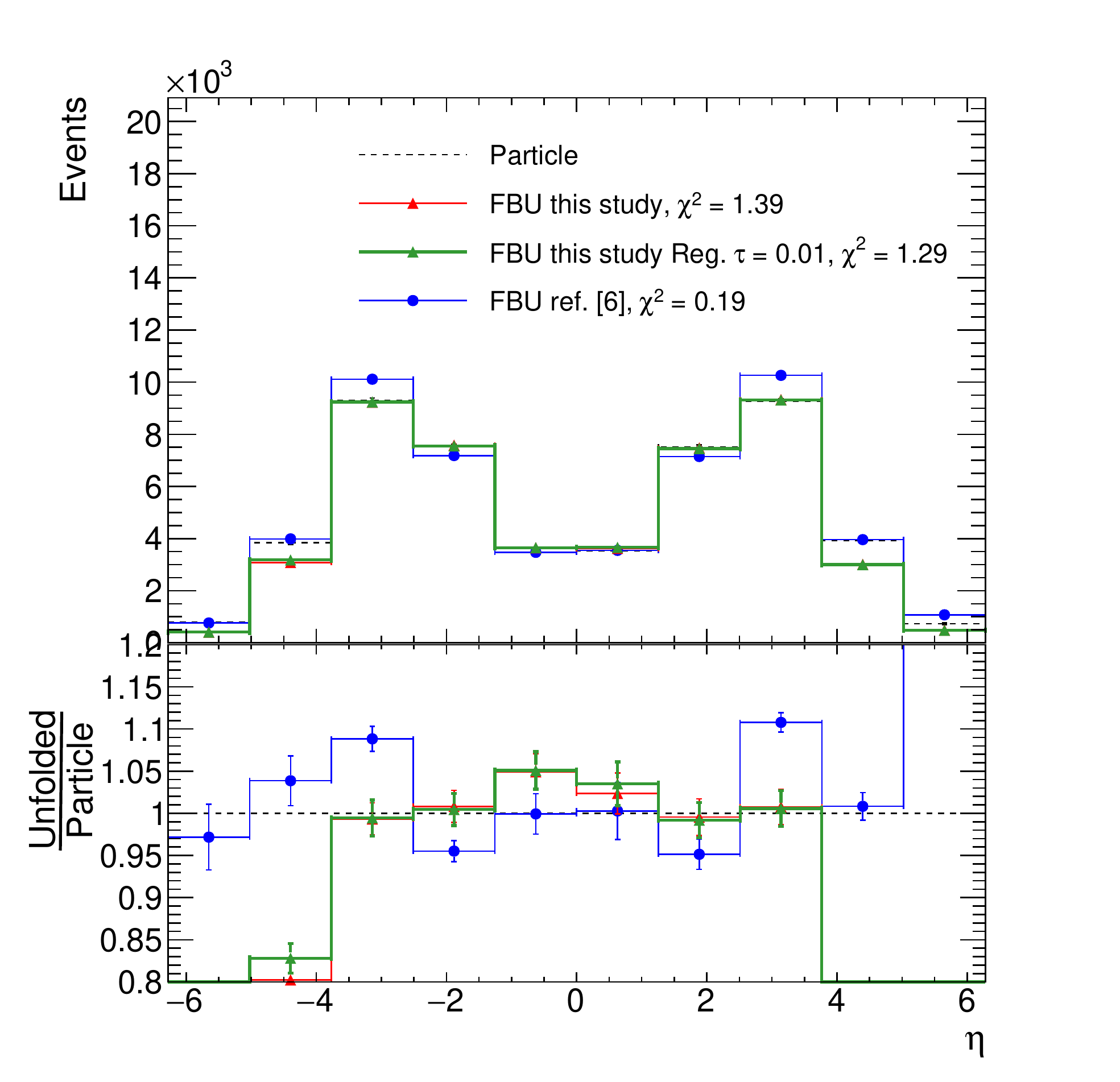}}
\subfloat[]
{\includegraphics[width=0.45\linewidth]{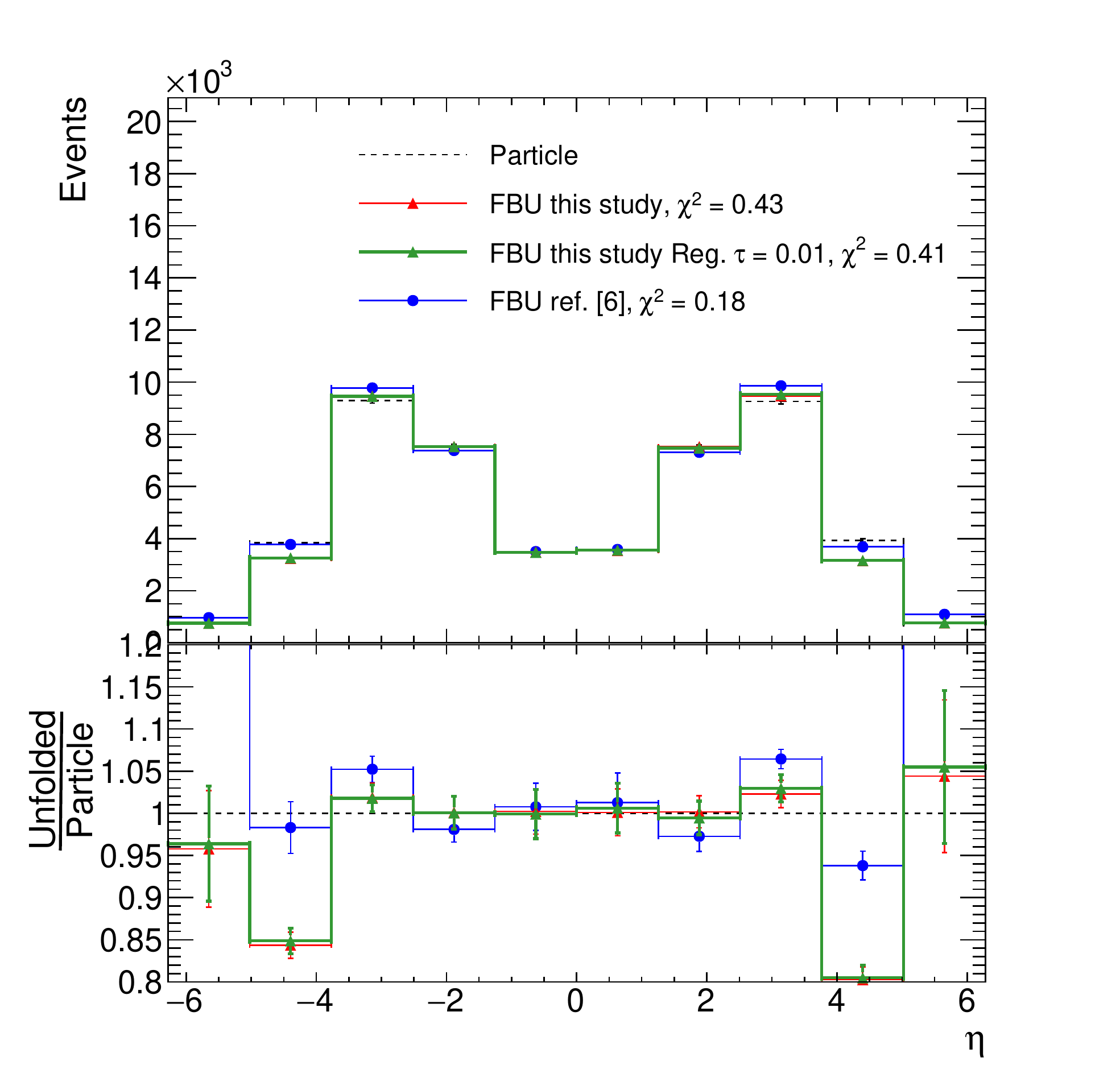}}
\caption{Figure 4: Unfolded pseudorapidity spectrum of the top quark pair. The green line is the unfolded spectrum with the use of regularization with parameter $\tau = 0.01$ in our implementation. The red line is the unfolded spectrum using our implementation without regularization. The blue line is the unfolded spectrum with use of the FBU unfolding code~\cite{Gerbaudo} using the mean of the posteriors. Plot (a) shows unfolded spectra after one iteration and plot (b) after two iterations.}
\label{fig:res}
\end{figure}

As a result the unfolded spectra of the top quarks pair pseudorapidity are presented after one and two iteration. The significant improvement between iterations is obvious from the decrease of the $\chi^2/\text{NDF}$ between each unfolded spectrum and the particle-level spectrum value in the plots (a) and (b) in the Figure \ref{fig:res}.\par Regularized results with the regularization strength $\tau = 0.01 $ are consistent within the statistical uncertainty with the unfolded spectra without using regularization (Figure \ref{fig:res}). \par



\section{Conclusion}
An iterative method was designed to improve unfolding results and to speed up the computation time. \par
Fully Bayesian Unfolding with regularization can be helpful in a specific kind of spectra. In the case of our implementation, the result of the pseudorapidity spectrum of the top anti-top quark pair shows that applying regularization we can obtain a better agreement in the first iteration.\par
However, applying the second iteration the differences between regularized and non-regularized spectrum vanishes.\par
In further analysis, the author would like to implement an algorithm to estimate the regularization strength $\tau$.
\vspace{-6pt}
\Acknowledgements
The author gratefully acknowledges the support from the project IGA\_PrF\_2019\_008 of Palacky University as well as grants of MSMT, Czech Republic, GACR 19-21484S and LTT-17018.

\end{document}